\documentclass[12pt]{iopart}
\newcommand{\be}{\begin{equation}}
\newcommand{\ee}{\end{equation}}
\newcommand{\bea}{\begin{eqnarray}}
\newcommand{\eea}{\end{eqnarray}}

\vspace{.5in}
\usepackage{epsfig}
\begin{document}
\title{Staggered and short period solutions of the Saturable Discrete
Nonlinear Schr{\"o}dinger Equation}
\author{Avinash Khare$^1$, Kim~{\O}. Rasmussen$^2$, Mogens R. Samuelsen$^3$, and Avadh Saxena$^2$}
\address{$^1$Institute of Physics, Bhubaneswar, Orissa 751005, India}
%\author{Kim~{\O}. Rasmussen}
\address{$^2$Theoretical Division, Los Alamos National Laboratory,
Los Alamos, New Mexico, 87545, USA}
%\author{Mogens R. Samuelsen}
\address{$^3$Department of Physics, The Technical University of Denmark, DK-2800 Kgs. Lyngby, Denmark}
%\author{Avadh Saxena}
%\affiliation{Theoretical Division, Los Alamos National Laboratory,
%Los Alamos, New Mexico, 87545, USA}
\date{\today}


\begin{abstract}
We point out that the nonlinear Schr{\"o}dinger lattice with a saturable 
nonlinearity also admits staggered periodic as well as localized pulse-like
solutions. Further, the same model also admits solutions with a short period.
We examine the stability of these solutions and find that the staggered
as well as the short period solutions are stable in most cases. We also show 
that the effective Peierls-Nabarro barrier for the pulse-like soliton 
solutions is zero.
\end{abstract}
\maketitle

The saturable discrete nonlinear Schr\"odinger equation is increasingly 
finding applications in various physical situations.  Most notably it serves as a 
model for optical pulse propagation in optically modulated photorefractive media \cite{Flei}, 
and in this context the pulse dynamics it describes have been intensely studied \cite{fitrakis,vicencio,cuevas}.
In addition to its important role for such applications the 
saturable discrete nonlinear Schr\"odinger equation is also of interest from 
a purely nonlinear science viewpoint \cite{melvin_1,oxtoby,melvin_2}. This 
interest arises because the saturable discrete nonlinear Schr\"odinger equation 
has been demonstrated \cite{had} to admit onsite and intersite soliton solutions, which 
have the same energy. This is contrasted by the standard cubic nonlinear Schr{\"o}dinger lattice where the 
onsite solution always has lower energy than the intersite solution. This phenomenon
have often been characterized in terms of a so-called Peierls-Nabarro (PN) barrier, 
which is the energy difference between these two distinct solutions. The particular 
feature of the saturable discrete nonlinear Schr\"odinger equation is thus that 
it allows the PN barrier to change sign and specifically vanish for certain solutions. 
The vanishing of the PN barrier have been associated with the ability of these solutions to 
translate undisturbed through the lattice, which is impossible in the cubic discrete 
nonlinear Schr\"odinger equation. Here we derive analytical solutions to the 
saturable discrete nonlinear Schr\"odinger equation
and demonstrate that the localized soliton solutions have a zero PN barrier.

%Discrete vector solitons \cite{fitrakis} and soliton mobility 
%in waveguide arrays with saturable nonlinearity \cite{vicencio} were also 
%studied in detail.    

Recently, we obtained \cite{krss} two different temporally and spatially 
periodic solutions to the saturable equation\cite{standard}
\be\label{1}
i\dot{\psi}_n + (\psi_{n+1}+\psi_{n-1}-2\psi_n)+
\frac{\nu|\psi_n|^2}{1+\mu|\psi_n|^2}\psi_n=0\,, 
\ee
where $\psi_n$ is a complex valued `wavefunction' at site $n$, while $\nu$ 
and $\mu$ are real parameters.  In particular, the first solution is 
\be\label{2}
\psi_n^I=\frac{1}{\sqrt{\mu}}\frac{\mbox{sn}(\beta,m)}{\mbox{cn}(\beta,m)}
\mbox{dn}([n+c]\beta,m)
\exp \left (-i[\omega t+\delta] \right),
\ee
where the modulus of the elliptic functions $m$ must be chosen such that
\be\label{3}
2-\omega=\frac{\nu}{\mu}=\frac{2\mbox{dn}(\beta,m)}{\mbox{cn}^2(\beta,m)},
~~~\beta=\frac{2K(m)}{N_p}\,,
\ee
and $c$ and $\delta$ are arbitrary constants. We only need to consider c between
0 and $\frac{1}{2}$ (half the lattice spacing).  Here $K(m)$ denotes 
the complete elliptic integral of the first kind \cite{stegun}. The second 
solution is  
\be\label{4}
\psi_n^{II}=\sqrt{\frac{m}{\mu}}\frac{\mbox{sn}(\beta,m)}{\mbox{dn}(\beta,m)}
\mbox{cn}([n+c]\beta,m)
\exp \left (-i[\omega t+\delta] \right)\,,
\ee
where the modulus $m$ is now determined such that
\be\label{5}
2-\omega=\frac{\nu}{\mu}=\frac{2\mbox{cn}(\beta,m)}{\mbox{dn}^2(\beta,m)},
~~~\beta=\frac{4K(m)}{N_p}\,.
\ee
The integer $N_p$ denotes the spatial period of the solutions. In the limit 
$N_p\rightarrow\infty$ ($m\rightarrow 1$), both the solutions $\psi_n^I$ and 
$\psi_n^{II}$ reduce to the same localized solution 
\be\label{6}
\psi_n^{III}=\frac{1}{\sqrt{\mu}}\frac{\sinh(\beta)}{\cosh([n+c]\beta)}
e^{-i[\omega t+\delta]},
~~(N_p\rightarrow\infty),
\ee
where $\beta$ is now given by
\be\label{7}
2-\omega = \frac{\nu}{\mu}=2\mbox{cosh}\beta\,. 
\ee
In Ref. \cite{krss}, we also developed the stability analysis and examined the linear stability of these solutions to
show that the solutions are linearly stable in most cases.

\begin{figure}
\begin{center}
\includegraphics[width=0.45\textwidth]{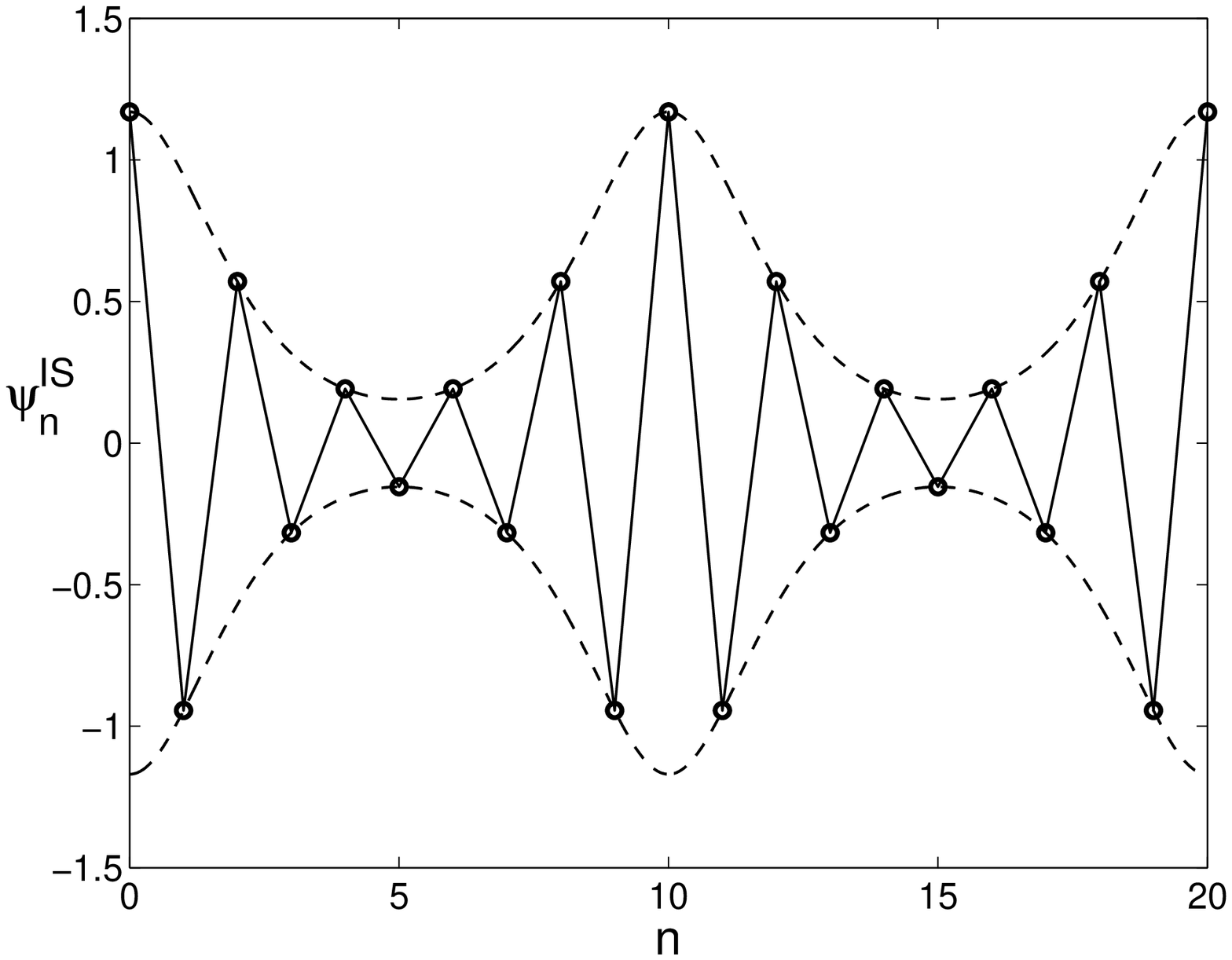}
\includegraphics[width=0.45\textwidth]{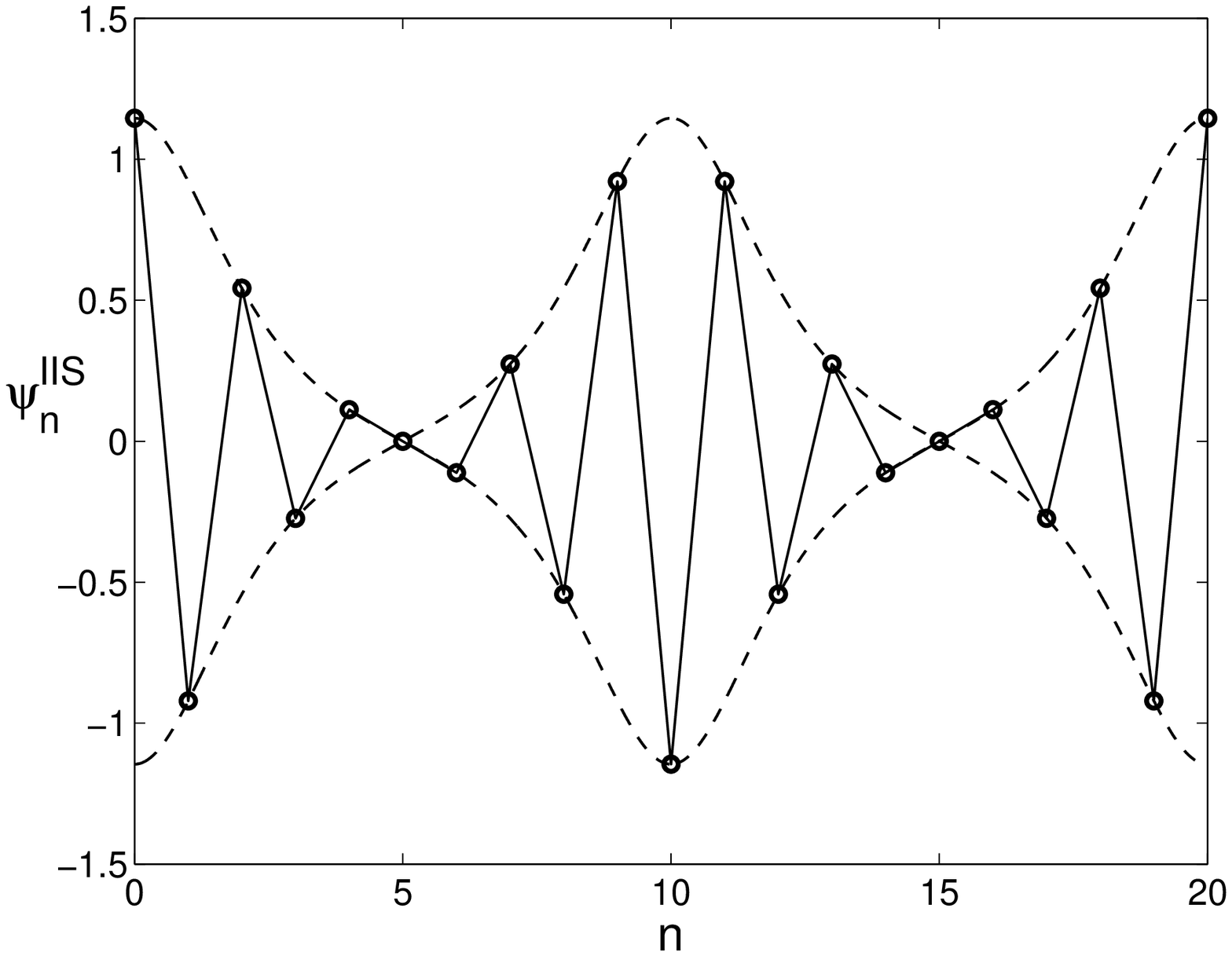}
\end{center}
 \caption{\label{fig:sol}Illustration of the exact solutions of two types. 
The parameters are: $\nu=-1$, $\mu=0.4$, $\omega=4.5$, and $c=t=\delta=0$. 
$N_p=10$ for $\psi_n^{IS}$ and 20 for $\psi_n^{IIS}$.  The dashed curves 
represent the solutions given by Eqs. (\ref{8}) and (\ref{10}) as if $n$ 
is a continuous variable. Lines are guides to the eye.}
 \end{figure}

The purpose of this note is to point out that the same model (\ref{1}) also
admits the corresponding staggered solutions. In particular, using the 
identities for the Jacobi elliptic functions \cite{khare}, it is easily
shown that the model admits the following solutions 
\be\label{8}
\psi_n^{IS}=(-1)^{n} \frac{1}{\sqrt{\mu}}\frac{\mbox{sn}(\beta,m)}
{\mbox{cn}(\beta,m)} \mbox{dn}([n+c]\beta,m)
\exp \left (-i[\omega t+\delta] \right)\,,
\ee
where the modulus $m$ must be chosen such that
\be\label{9}
\omega-2=-\frac{\nu}{\mu}=\frac{2\mbox{dn}(\beta,m)}{\mbox{cn}^2(\beta,m)},
~~~\beta=\frac{2K(m)}{N_p}\,.
\ee
\be\label{10}
\psi_n^{IIS}=(-1)^{n}\sqrt{\frac{m}{\mu}}\frac{\mbox{sn}(\beta,m)}
{\mbox{dn}(\beta,m)} \mbox{cn}([n+c]\beta,m)
\exp \left (-i[\omega t+\delta] \right)\,,
\ee
where the modulus $m$ is now determined such that
\be\label{11}
\omega-2=-\frac{\nu}{\mu}=\frac{2\mbox{cn}(\beta,m)}{\mbox{dn}^2(\beta,m)},
~~~\beta=\frac{4K(m)}{N_p}\,.
\ee
In the limit $N_p\rightarrow\infty$ ($m\rightarrow 1$),
both the solutions $\psi_n^{IS}$ and $\psi_n^{IIS}$ reduce to the same 
localized staggered solution 
\be\label{12}
\psi_n^{IIIS}=(-1)^{n} \frac{1}{\sqrt{\mu}}\frac{\sinh(\beta)}
{\cosh([n+c]\beta)} e^{-i[\omega t+\delta]},
~~~(N_p\rightarrow\infty)\,,
\ee
where $\beta$ is now given by
\be\label{13}
\omega-2 = -\frac{\nu}{\mu} =2\mbox{cosh}\beta\,.
\ee
As an illustration  we have plotted the exact solutions of the
type IS and IIS in Fig. \ref{fig:sol}. Here the period $N_p$ has to be even.
We have shown two periods for type IS and only one for type IIS.

There are, as expressed by Eqs. (\ref{9}), (\ref{11}), and 
(\ref{13}), stringent conditions on the parameters $\mu$ and $\nu$ 
for which these exact solutions exist. 
For example, while the nonstaggered solutions are only valid
for $\nu >0$ and hence $\omega <2$, the staggered solutions are valid only
if $\nu <0$ and hence $\omega >2$. In the case IS the limitation is
\be\label{caseIS}
0~(m=1) < -\frac{2\mu}{\nu} < \cos^2\left(\frac{\pi}{N_p}\right) ~(m=0)\,,
\ee
while in the case IIS the limitation is
\be\label{caseIIS}
0~(m=1) < -\frac{2\mu}{\nu} < \frac{1}{\cos(\frac{2\pi}{N_p})} ~(m=0)\,.
\ee
Similarly, the  solution $\psi_n^{IIIS}$ exists only when $-\frac{2\mu}{\nu}$ 
is close to zero (m=1).

We have also examined the linear stability of these solutions and find 
that the solutions are linearly stable in most cases. A single period $(N=N_p$i, where $N$ is the lattice size)
is always stable for both solutions IS and IIS. A type IIS solution with 
more than one period ($N=jN_p$, where $j$ is an integer larger than one) 
is also stable, while a type IS solution with more than one period is always 
unstable. Thus, the first example in Fig. \ref{fig:sol} is in fact unstable.

For the solution IIIS, expressions for both the power and the Hamiltonian  
are identical to those for the solution III and are given by Eqs. (13) and 
(14) of Ref. \cite{krss}.  Hence the PN barrier for the 
solutions III and IIIS is the same.  We would like to point out here that 
the calculation of PN barrier in I was not quite correct. In particular, 
since both power P and the Hamiltonian H are constants of motion, one must 
compute the energy difference between the solutions when $c=0$ and $c=1/2$ 
in such a way that the power P is {\it same} in both the cases.  On using 
the expressions for P and H as given by Eqs. (13) and (14) of I, we find 
that H for the solution III as well as IIIS is given by
\be\label{14}
H =-\frac{4\sinh(\beta)}{\mu}+\frac{2\beta \nu}{\mu^2}
+2\left(1-\frac{\nu}{2\mu}\right)P\,.
\ee
Note that H is in fact independent of $c$, i.e. contrary to our claim in 
Ref. \cite{krss}, the PN barrier is in fact zero for our solution III (and 
hence also for IIIS). 

Before completing this note, we would like to mention that the model 
(\ref{1}) also admits a few short period solutions.

Using the ansatz  
\be\label{3.2}
\psi_n(t)=\phi_n e^{-i(\omega t+\delta)}\,, 
\ee

in Eq. (\ref{1}) it is easily checked that the only possible short period 
solutions are to Eq. (\ref{1}) are:

\begin{enumerate}
\item Period 1 solution $\phi_n = (...,a,a,...)$ provided
\be\label{p.1}
\omega=-\frac{\nu a^2}{1+\mu a^2}.
\ee
\item Period 2 solution $\phi_n = (...,a,-a,...)$ provided
\be\label{p.2}
\omega=4-\frac{\nu a^2}{1+\mu a^2}.
\ee

\item Period 3 Solution  $\phi_n = (...,a,0,-a,...)$ provided
\be\label{p.3}
\omega=3-\frac{\nu a^2}{1+\mu a^2}.
\ee

\item Period 4 Solutions $\phi_n = (...,a,0,-a,0,...)$ and 
$(...,a,a,-a,-a,...)$ provided
\be\label{p.4}
\omega=2-\frac{\nu a^2}{1+\mu a^2}.
\ee

\item Period 6 Solution  $\phi_n = (...,a,a,0,-a,-a,0,...)$ provided
\be\label{p.6}
\omega=1-\frac{\nu a^2}{1+\mu a^2}.
\ee

\end{enumerate}
Applying the stability analysis developed in Ref. \cite{krss} we have 
examined the stability of these short period solutions and find that for 
a small nonlinearity ($|\nu|<2\mu$) they are all stable. The period 4 solution $(...,a,a,-a,-a,...)$ is always 
stable while all the other short period solutions
possess regions of instabilities at larger nonlinearity. For these low period solutions the 
stability matrices given by Eqs. (20) and (21) of Ref. \cite{krss} are simple and it is, for 
example, easy to see that the lowest non-zero eigenvalue, $\lambda_1(a,\nu)$, of 
the stability problem for the period 1 solutions is 
given by ($\mu =1$)
\be
\lambda_1(a,\nu)=a^4 +\left(2-\frac{2}{3}\nu\right)a^2+1.
\ee
Similarly, we have for the period 2 solution
\be
\lambda_2(a,\nu)=a^4 +\left(2+\frac{2}{3}\nu\right)a^2+1,
\ee
and the period 4 solution
\be
\lambda_4(a,\nu)=a^4 +(2-|\nu|)a^2+1.
\ee
These expressions correspond to the relevant curves in Fig. 2. 

It possible to derive similar expressions for the period 3 and period 6 
solutions but the expressions are more complicated and will be omitted here. 
\begin{figure}[h]
\begin{center}
\includegraphics[width=0.45\textwidth]{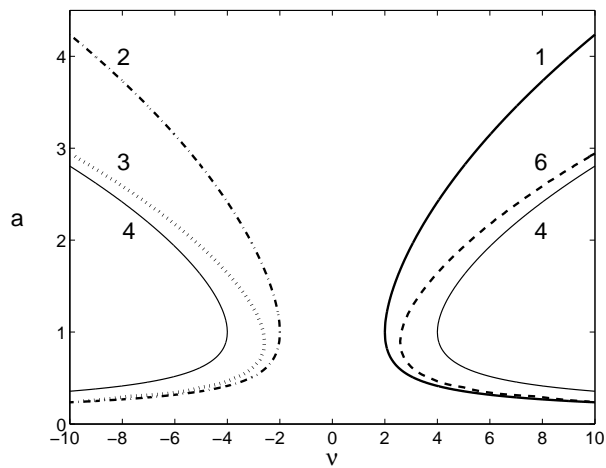}
\end{center}
 \caption{\label{stabil}Regions of stability for the short period
solutions to Eq. (\ref{1}) for $\mu=1$. Period 1: thick full curve, period 2:
dashed-dotted curve, period 3: long-dashed curve, period 4: thin full curve, and
period 6: short-dashed curve. The instability occurs in the parameter region encompassed by the 
respective curves. 
}
\end{figure}
Clearly the $p$ period solutions are unstable for the parameter values where 
$\lambda_p(a,\nu)<0$, and we have illustrated these regions in Fig. 
\ref{stabil}. Figure \ref{stabil} shows the curves in the $(a,\nu)$-plane 
where $\lambda_p(a,\nu)=0$ so that the instability occurs in the regions 
that are encompassed by the respective curves. 
%The first 
%period 4 solution is always stable, while the other solutions become
%unstable for larger nonlinearity depending on $a$ as shown in Fig. 2. Note 
%that period 2 and 3 become unstable only for negative values of $\nu$, and  
%period 1 and 6 only for positive values of $\nu$, while period 4 becomes 
%unstable for both positive and negative values. 
A symmetry is apparent in this stability diagram and it is easy to realize 
that this symmetry arises from the fact that the transformation $(\nu,\phi_n) 
\rightarrow (-\nu,(-1)^n \phi_n)$ establishes the following connection 
between the 
%the figure, a symmetry which follows from the fact that a transformation 
%multiplying one short period solution by $(-1)^n$ brings it into another 
short period solutions: 1 $\leftrightarrow$ 2, 3 $\leftrightarrow$ 6, and 4 
$\leftrightarrow$ 4.  

In conclusion, we have obtained staggered as well as short period solutions 
of the saturable discrete nonlinear Schr\"odinger equation.  We also studied 
the linear stability and found the solutions to be stable in certain 
parameter ranges.  Finally, we found that the Peierls-Nabarro barrier for the 
pulse solutions is zero.  Our results are relevant to optical soliton pulse 
propagation in waveguides and photorefractive media \cite{Flei}.  

Research at Los Alamos National Laboratory is carried out under the auspices 
of the National Nuclear Security Administration of the U.S. Department of 
Energy under Contract No. DE-AC52-06NA25396.


\begin{thebibliography}{10}
\bibitem{standard} Note that rewritting $\frac{\nu|\psi_n|^2}{1+\mu|\psi_n|^2}\psi_n=\frac{\nu}{\mu}(1-\frac{1}{1+\mu|\psi_n|^2})\psi_n$ 
and the notation change $\frac{\nu}{\mu} \rightarrow -\nu$ followed by
the gauge transformation $\phi_n \rightarrow \phi_n\exp(-\nu t)$ render the equantion in a perhaps more often used form \cite{melvin_2}.
\end{thebibliography}
\end{document}